# EFFICIENT HARDWARE DESIGN AND IMPLEMENTATION OF ENCRYPTED MIPS PROCESSOR

Kirat Pal Singh, Centre for Development of Advanced Computing (C-DAC), Mohali, Punjab, India
Kirat_addiwal@yahoo.com
Dilip Kumar, Centre for Development of Advanced Computing (C-DAC), Mohali, Punjab, India
dilip.k78@gmail.com

**ABSTRACT**

*The paper describes the design and hardware implementation of 32-bit encrypted MIPS processor based on MIPS pipeline architecture. The organization of pipeline stages in such a way that pipeline can be clocked at high frequency. Encryption and Decryption blocks of data encryption standard (DES) cryptosystem and dependency among themselves are explained in detail with the help of block diagram. In order to increase the processor functionality and performance, especially for security applications we include three new instructions 32-bit LKLW, LKUW and CRYPT. The design has been synthesized at 40nm process technology targeting using Xilinx Virtex-6 device. The encrypted MIPS pipeline processor can work at 218MHz at synthesis level and 744MHz at simulation level.*

**Keywords:** *Processor, ALU, Register file, pipeline, DES, Throughput*

## INTRODUCTION

The MIPS is simply known as Millions of instructions per second and is one of the best RISC (Reduced Instruction Set Computer) processor ever designed. High speed MIPS processor possessed Pipeline architecture for speed up processing, increase the frequency and performance of the processor. A MIPS based RISC processor was described in (*D. A. Patterson, et al., 2005*). It consist of basic five stages of pipelining that are Instruction Fetch, Instruction Decode, Instruction Execution, Memory access, write back. These five pipeline stages generate 5 clock cycles processing delay and several Hazard during the operation (*Zulkifli, et al., 2009*). These pipelining Hazard are eliminates by inserting NOP (No Operation Performed) instruction which generate some delays for the proper execution of instruction (*D. A. Patterson, et al., 2005*). The pipelining Hazards are of three type's data, structural and control hazard. These hazards are handled in the MIPS processor by the implementation of forwarding unit, Pre-fetching or Hazard detection unit, branch and jump prediction unit (*Zulkifli, et al., 2009*). Forwarding unit is used for preventing data hazards which detects the dependencies and forward the required data from the running instruction to the dependent instructions (*Pejman lotfi, et al., 2011*). Stall are occurred in the pipelined architecture when the consecutive instruction uses the same operand of the instruction and that require more clock cycles for execution and reduces performance. To overcome this situation, instruction pre-fetching unit is used which reduces the stalls and improve performance. The control hazard are occurs when a branch prediction is mistaken or in general, when the system has no mechanism for handling the control hazards (*Pejman lotfi, et al., 2011*). The control hazard is handled by two mechanisms: Flush mechanism and Delayed jump mechanism. The branch and jump prediction unit uses these two mechanisms for preventing control hazards. The flush mechanism runs instruction after a branch and flushes the pipe after the misprediction (*Pejman lotfi, et al., 2011*). Frequent flushing may increase the clock cycles and reduce performance. In the delayed jump mechanism, to handle the control hazard is to fill the pipe after the jump instruction with specific numbers of NOP's (*Pejman lotfi, et al., 2011*). The branch and jump prediction unit placement in the pipelining architecture may affect the critical or longest path. To detecting the longest path and improving the hardware that resulting minimum clock period and is the standard method of increasing the performance of the processor. The MIPS architecture employs a wide range of applications. The architecture remains the same for all MIPS based processors while the implementations may differ (*Gautham P, et al., 2009*). The proposed design has the feature of 32-bit asymmetric and symmetric cryptography system as a security application. There is a 16- bit RSA cryptography MIPS cryptosystem have been previously designed (*Zulkifli, et al., 2009*). There is the small

Efficient Hardware Design and Implementation of Encrypted MIPS Processor

adjustments and minor improvement in the MIPS pipelined architecture design to protect data transmission over insecure medium using authenticating devices such as data encryption standard [DES], 3DES and Advanced Encryption Standard (AES) (*Pravin B. ghewari, et al., 2010*). These cryptographic devices use an identical key for the receiver side and sender side. Our design mainly includes the symmetric cryptosystem into MIPS pipeline stages. That is suitable to encrypt large amount data with high speed.

**SYSTEM ARCHITECTURE**

The global architecture of a conventional pipelined processor employs 5 basic pipeline stages that are Instruction Fetch (IF), Instruction Decode [ID], Instruction Execution (EXE), Memory Access (MEM), Write Back (WB). These pipeline stages operate concurrently, using synchronization signals: Clock and Reset. MIPS architecture employs three different Instruction format: R-Type Instruction, I-Type Instruction, and J-Type Instruction (*Pravin B. ghewari, et al., 2010*).

The simple MIPS processor has various sub-blocks inside the pipeline stages shown in *Figure 1*. The instruction fetch unit contains Program counter (PC), and Instruction Memory. The function of Instruction fetch unit is to obtain an instruction from the instruction memory using the current value of PC and increment PC value for next instruction and placed that value to IF register. The Instruction decode unit contain Instruction decoder, register file, sign extender, Control unit. The function of this unit is to obtain 32-bit instruction from IF register to index the register file and register data. This unit also sign extends 16-bit instruction to 32-bit instruction. The data written to the register file at the proper address are from write back stage. The control unit sends various control signals to other pipeline stage. The execution unit contain arithmetic logic unit (ALU) which performs the arithmetic and logical operation determined by the ALU operation signal which comes from control unit. The memory access unit contain data memory which load and store instruction data.

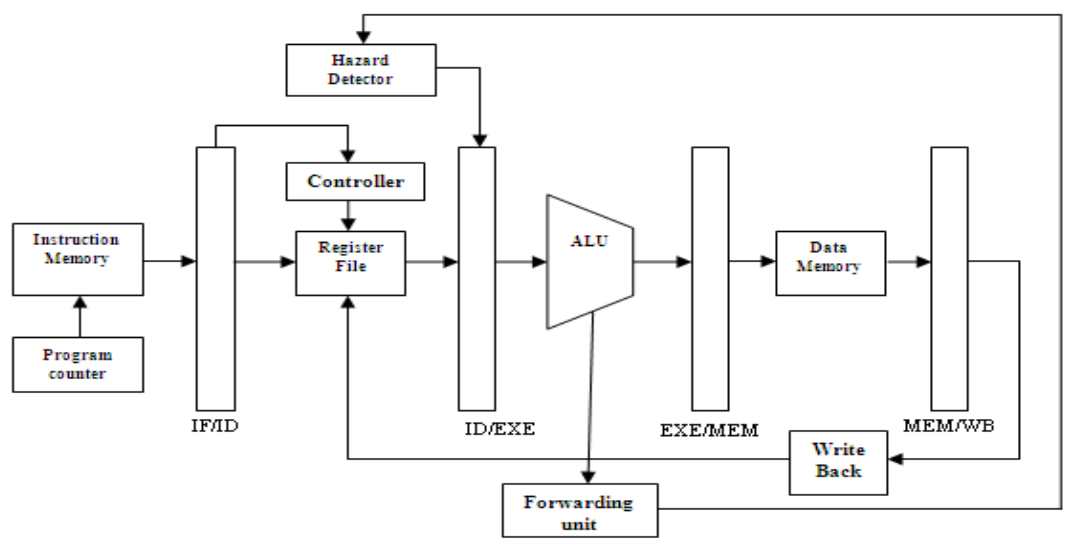

*Figure 1: Pipelined MIPS processor 'Architecture'*

The Encrypted MIPS processor is generally based on MIPS architecture. The pipelined MIPS architecture is modified in such a way that it executes encrypted instruction. *Figure 2* shows the block diagram of encrypted MIPS processor. To modify MIPS processor for encryption we just insert the cryptography module such as Data Encryption Standard (DES), Triple Data Encryption Standard (T-DES), Advanced Encryption Standard (AES) etc. to the pipeline stage. Only single cryptographic module is used in same hardware implementation. The instruction fetch unit of encrypted MIPS contains Program Counter (PC), instruction memory, decryption core and MUX. The instruction memory read address from PC and store instruction value at the particular address that points by the PC. Instruction memory sends encrypted instruction to MUX and decryption core. The decryption core give decrypted instructions and further send to the MUX and output of MUX is fed to the IF register. The MUX control signal comes from control unit.

Efficient Hardware Design and Implementation of Encrypted MIPS Processor

The instruction decode unit contain register file and key register. Key register store the key data of encryption/decryption core. Key address and key data comes from write back stage. Once the key data to be stored into register file it will remain same for all program instruction execution. The control unit provides various control signals to other stages. The execute unit executes the register file output data and perform the particular operation determined by the ALU. The ALU output data send to EXE register. The memory access unit contains encryption core, decryption core, data memory, MUX, DEMUX. The second register data from register file fed to the encryption core and also to MUX. Here the crypt signal enable/disable encryption operation. The read/write signal of data memory describes whether reading/writing operation is done. Output of data memory pass through DEMUX and its one output go to decryption core and other goes to MEM register. Here the unencrypted memory data and decrypted data temporarily store to MEM register. The MEM output fed to write back data MUX and according to control signal the output of MUX goes to register file.

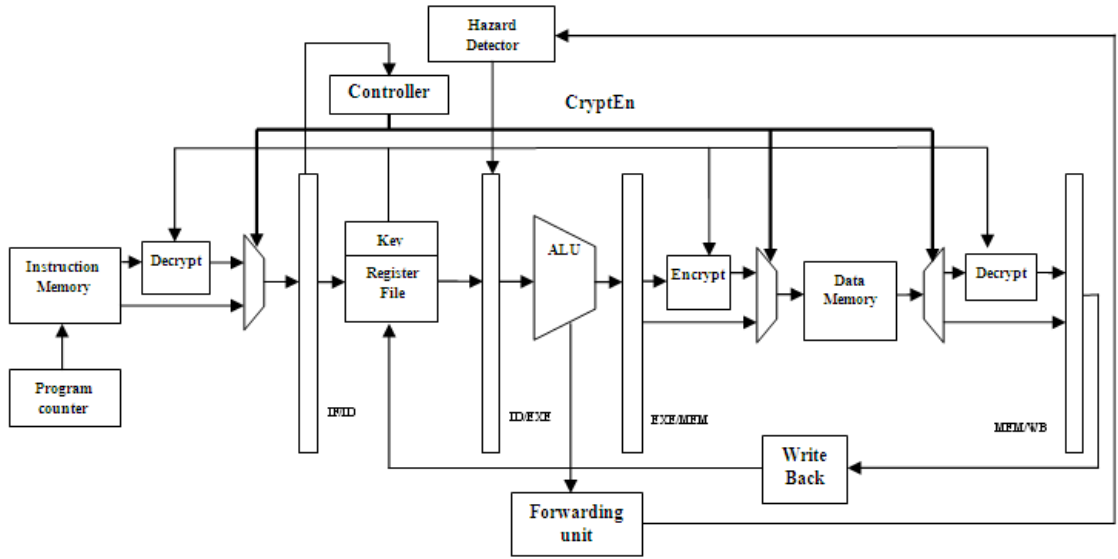

*Figure 2: Modified Encrypted MIPS processor 'Hardware Implementation*

**MIPS Instruction set**

The MIPS instruction set is straightforward like other RISC designs. MIPS are a load/store architecture, which means that only load and store instructions access memory. Other instructions can only operate on values in registers (*Rashmi S. Keote, et. al., 2011*). Generally, the MIPS instructions can be broken into three classes: the memory-reference instructions, the arithmetic- logical instructions, and the branch instructions. Also, there are three different instructions formats in MIPS architecture: R-Type instructions, I-Type instructions, and J-Type instructions as shown in *Figure 3*.

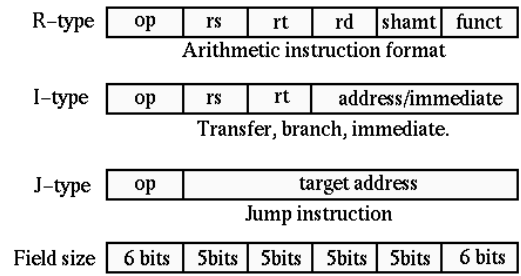

*Figure 3: MIPS Instruction Types*

The MIPS instruction field is described in *Table 1*. There are three more new instructions that supports encrypted operation. These instructions are load key upper word (LKUW), load key lower word (LKLW) and encryption mode (CRYPT). These instructions randomly used opcodes in the hardware implementation. LKLW and LKUW come under I-type instruction and variant of load word (LW). These two instruction need not to specify a destination address in the assembly code. CRYPT instruction comes under J-type instruction and instead of address, only single argument i.e. Boolean value is to be assigned. This indicates enable/disable encryption process. Any nonzero value enables the encryption process and zero value disables the encryption process.

*Table 1: MIPS Instruction Field*

| Field | Description |
| --- | --- |
| *Op[31-26]* | is a 6-bit operation code |
| *RS[25-21]* | is a 5-bit source register specifier |
| *RT[20-16]* | is a 5-bit target(source/destination) register or branch condition |
| *Immediate[15-0]* | is a 16-bit immediate, branch displacement or address displacement |
| *Target[25-0]* | is a 26-bit jump target address |
| *RD[15-11]* | is a 5-bit destination register specifier |
| *Shamt[10-6]* | is a 5-bit shift amount |
| *Funct[5-0]* | is a 6-bit function field |

**IMPLEMENTATION METHODOLOGY**

The complete hardware implementation of design process is shown in Fig. 4. We use the DES Crypto core which supports both encryption and decryption (*Saeid Taherkhani, et. al., 2010*). DES core has a 64-bit plaintext input, 64-bit key input, start signal, encryption/decryption enable signal, and 64-bit cipher text output. The 32-bit encrypted processor pack two 32-bit MIPS instruction into a single 64-bit instruction block for DES encryption/decryption process and breakout each individual instruction into the hardware. In this processor, there is some unencrypted instruction stored in data memory as a zero-padded 64-bit word. The program counter increment by 8 instead of 4 due to loading of 64-bit instructions. Both data and instruction memory reads 64-bit instruction at a time.

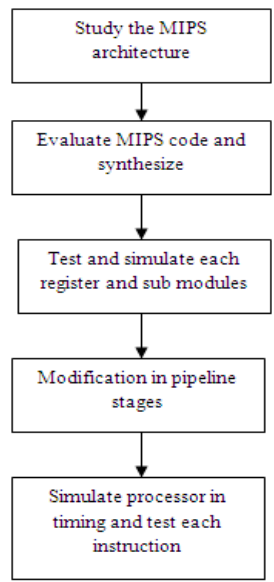

*Figure 4: Design Methodology of encrypted MIPS processor*

Before fetching the encrypted instruction the key loaded from memory is to be done and there may be a dependency of instruction takes place that causes hazard. So to overcome this dependency we may use NOP instruction that deactivates the control signal of current instructions. The forwarding unit also modifying to give the load key instruction. The NOP's are sufficient to insert between load key and CRYPT instruction. This is explained by an example.

```
addi $r1, $r0, 104
lklw 0($r1)
addi $r1, $r1, 8
lkuw 0($r1)
nop
nop
crypt 1
addi $r1, $r0, 7
add  $r2, $r0, $r0
addi $r3, $r0, 0
addi $r4, $r0, 0
Loop:   add  $r5, $r2, $r2
add  $r5, $r5, $r5
add  $r5, $r5, $r5
add  $r5, $r5, $r3
lw   $r6, 0($r5)
add  $r4, $r4, $r6
addi $r2, $r2, 1
slt  $r7, $r2, $r1
j    Loop
Exit:   sw   $r4, 56($r0)
```

*Figure 5: Pseudo code of encrypted MIPS Processor*

*Figure 5* shows the example of encrypted MIPS pipeline processor in assembly. The first instruction loads the base address for the key. Second instruction loads the lower word of the key at same register in upper instruction. Third instruction increment the base address of the key. Fourth instruction loads the upper word of the key. Next two instruction are NOP which indicates delay of two clock cycles for key to be loaded. CRYPT 1 instruction enables encryption process. Further, Next instructions are the simple MIPS program instructions that gives the output data which are stored in memory location 56. All these MIPS instructions after CRYPT are encrypted instructions.

**IMPLEMENTATION RESULTS**

The complete pipeline processor stages are modelled in VHDL. The syntax of the RTL design is checked using Xilinx tool. For functional verification of the design the MIPS processor is modelled in Hardware descriptive language. The design is verified both at a block level and top level. Test cases for the block level are generated in VHDL by both directed and random way. Table 2 shows the corresponding symbol and an architectural body in the RTL view. For top level verification assembly program are written and the corresponding hex code from the assembler is fed to both RTL design and model the checker module captures and compares the signal from both the model and display the message form mismatching of digital values.
The complete design along with all timing constraints, area utilization and optimization options are described using synthesis report. The design has been synthesized targeting 40nm triple oxide process technology using Xilinx FPGA Virtex-6 (xc6vlx240t-3ff1156) device. The Virtex family is the latest and fastest FPGA which aims to provide up to 15% lower dynamic and static power and 15% improved performance than the previous generation. It is obvious that there is a trade-off between maximum clock frequency and area utilization (number of slices LUT's) because the basic programmable part of FPGA is the slice that contain four LUTs (look up table) and eight Flip flops. Some of the slice can use their LUT's as distributed RAM.

Efficient Hardware Design and Implementation of Encrypted MIPS Processor

**SIMULATION RESULTS**

*Figure 6* shows the default input encrypted instruction memory contents (imemcontents) inside the instruction memory. These encrypted imemcontents values allocate starting address of 0 to 176 memory locations.

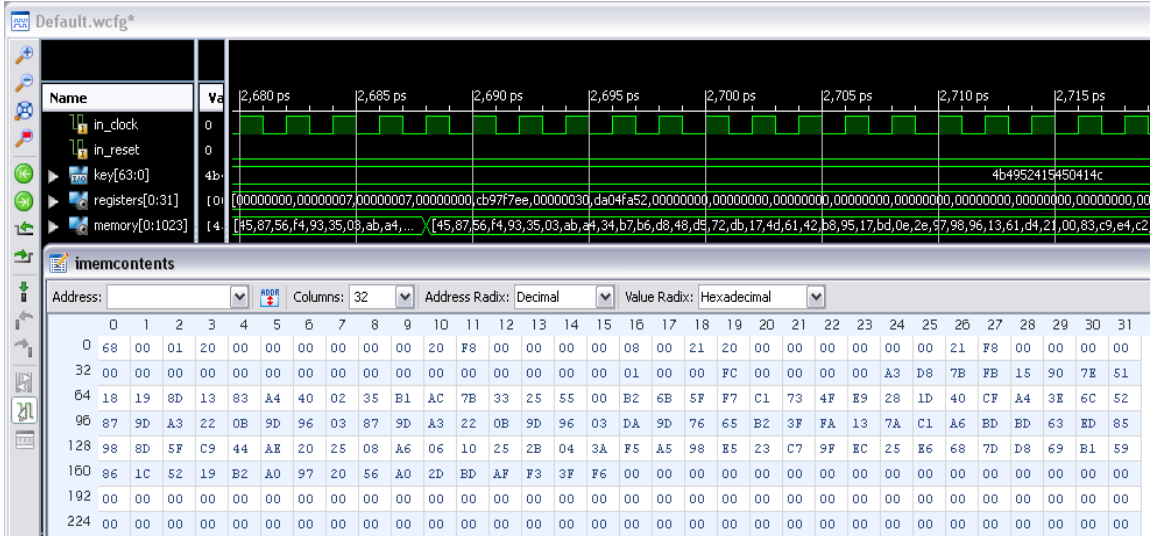

*Figure 6: Encrypted Instruction Inside Instruction Memory*

*Figure 7* shows the default input decrypted data memory contents (dmemcontents) inside the data memory and allocate memory address of 0 to 55 locations. The 64-bit unencrypted key value i.e. KIRATPAL (a string of 8 ASCII characters) in terms of Hex value i.e. 4b4952415450414c inside the data memory at starting address of 104 to 119 locations.

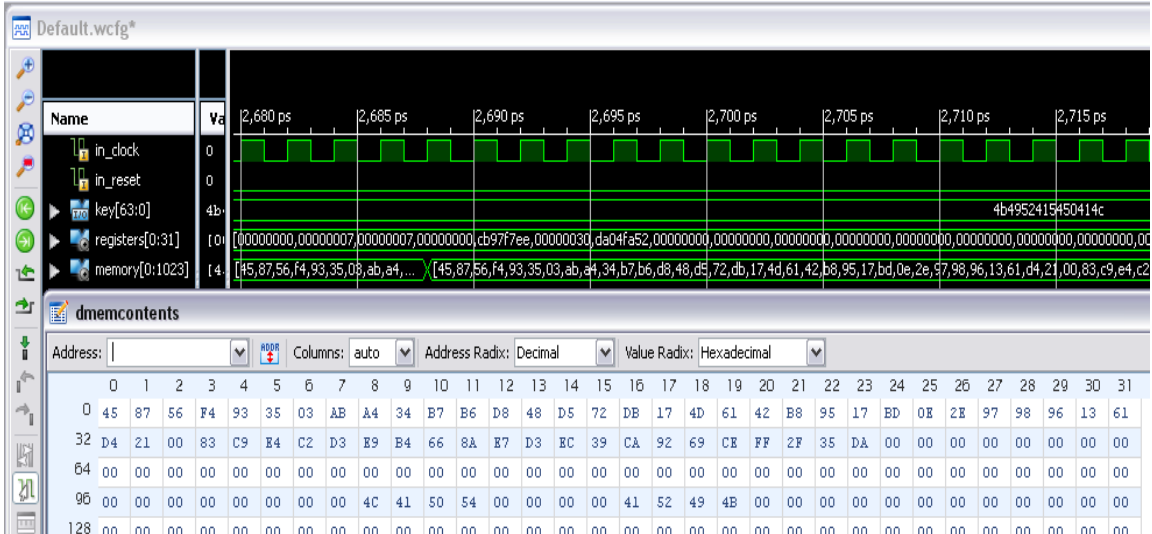

*Figure 7: Decrypted Instructions and Key Values Inside Data Memory*

*Figure 8* shows the resultant waveform generated by the 32-bit encrypted MIPS processor. The input is clock of 2ps time period, active high reset which initializes all processor subunit to zero. After clock period 2ps, active low reset all encrypted instructions are loaded and execute. The input is clock of 2688ps, active low reset, and the resultant encrypted output is obtained as cipher data.

Efficient Hardware Design and Implementation of Encrypted MIPS Processor

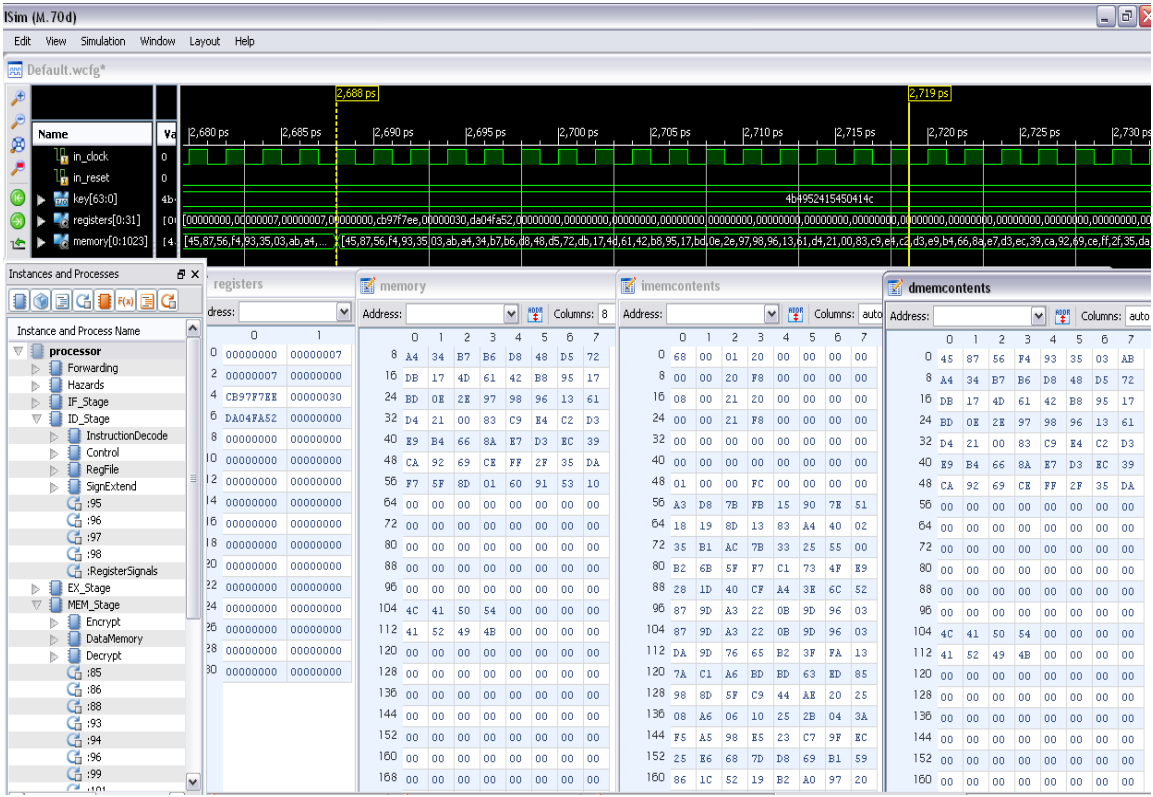

*Figure 8: Resulant Waveform of Encrypted Instruction*

*Figure 9* shows the output register values at 2688ps after executed all encrypted instruction from instruction memory. The plain text (input data) is stored at register (*Gautham P, Parthasarathy , et al, 2009)* .

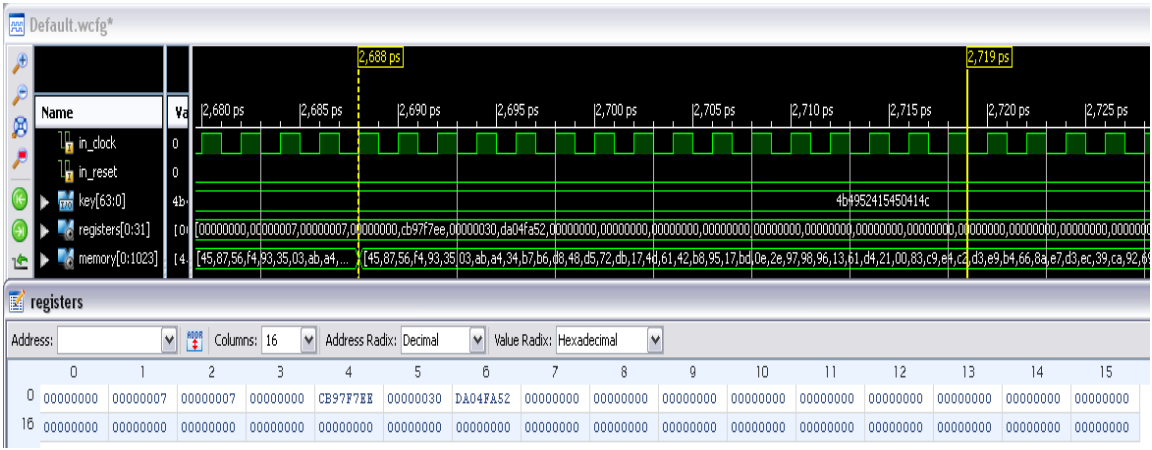

*Figure 9: Output Register Values at 2688ps*

*Figure 10* shows that output cipher data value is stored at memory location address of 56 (decimal). Before 2688ps the memory location is empty as shown in this figure. The resultant value is obtained after 2688ps as shown in above *Figure 8*. The lower byte of encrypted cipher data is stored at memory location of 56 (decimal) and upper byte is stored at 63(decimal). Hence, total 64-bit cipher data value is obtained.

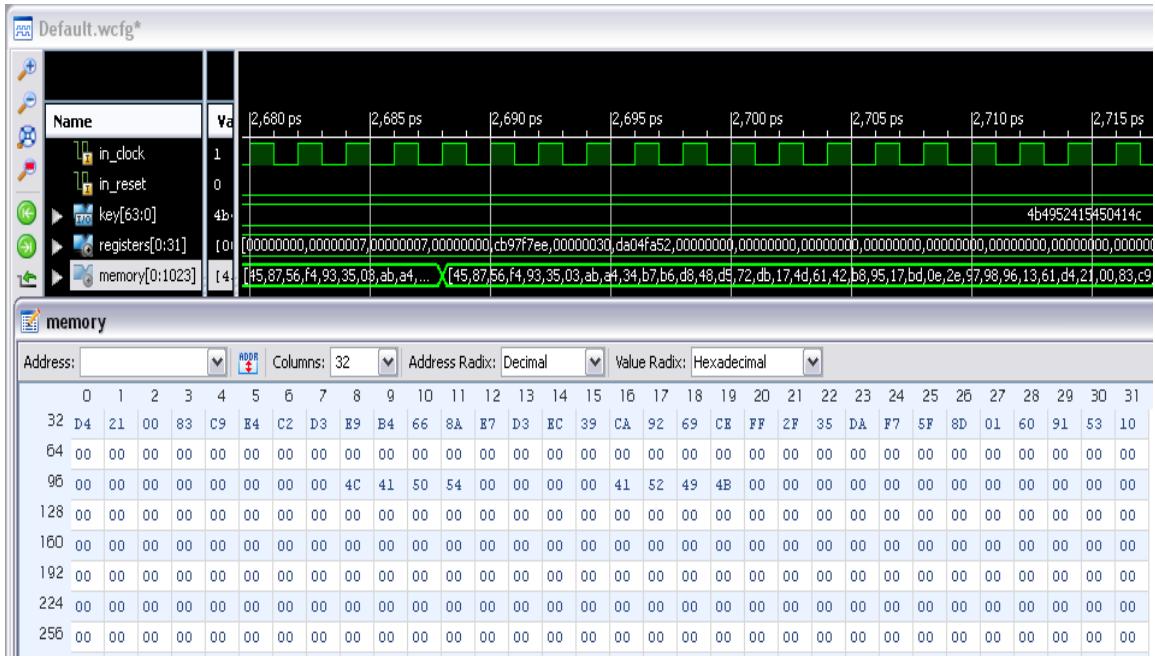

*Figure 10: Output Cipher Data Stored Inside Memory Address of 56(decimal) at 2688ps*

The encrypted MIPS processor found the correct value for sum of array i.e. cba767ee, which is stored in register (*Gautham P, Parthasarathy , et al, 2009*)
Input/plain text – cb97f7ee
Key – 4b4952415450414c
Output/cipher text – 10539160018d5ff7

**Verification and Synthesis**

The design is verified both at block level and top level. As system verification, we successfully execute encrypted MIPS program and DES encryption. Test case for block level is generated in VHDL by both directed and random way. Table 2 sows the corresponding symbol and an architectural body in the RTL view.

*Table 2: Subunits used in the Encrypted MIPS Processor*

| Stage | Subunits |
| --- | --- |
| *IF* | InstructionMemory, DES (Decrypt) |
| *ID* | InstructionDecode, Control, RegFile, SignExtend |
| *EXE* | ALU |
| *MEM* | DES (Encrypt), DataMemory, DES (Decrypt) |
| *WB* | Writeback |
| *Forwarding* | ForwardingUNIT |
| *Hazards* | HazardDetectionUnit |

The synthesis and mapping result of encrypted MIPS pipeline processor design are summarized in *Table 3*. The speed performance of the processor is affected by hardware (i.e. clock rate), instruction set, and compiler. The timing report of encrypted MIPS processor shows 4.584ns clock period at synthesis level and 1.343ns clock period at simulation level.

*Table 3: Results of FPGA Implementation of Encrypted MIPS Processor*

| Target FPGA Device | Virtex-6 (XC6vlx240t-3ff1156) |
|---|---|
| Process Technology | 40nm |
| Optimization Goal | Speed |
| Max. operating frequency (hardware) | 218MHz (synthesis level) |
| Max. operating frequency (software) | 744MHz (simulation level) |
| Number of slice registers | 10405 |
| Number of slice LUT's | 66072 |
| Number of fully used LUT flip flop pairs | 9305 |
| Number of bonded IOB's | 598 |
| Instructions throughput | 19Mbits/sec |

**CONCLUSION**

We proposed the efficient hardware architecture design and implementation of 32-bit encrypted MIPS processor. Which executes encrypted instructions, read and decrypt encrypted data from memory unit and write encrypted data back to memory. The processor uses the symmetric block cipher that can process data block of length 64-bits plain text, 64-bits key and 64-bit cipher data. The design has been modeled in VHDL and functional verification policies adopted for it. Optimization and synthesis of design is carried out at latest and fastest FPGA Viretx-6 device that improves performance. Each program instructions are tested with some of vectors provided by MIPS. We conclude that system implementation reach maximum frequency of 218MHz after synthesizing at 40nm process technology and 744MHz at simulation level.

- D. A. Patterson and J. L. Hennessy (2005), "Computer Organization and Design", the hardware/Software Interface.Morgan Kaufmann.
- Zulkifli, Yudhanto, Soetharyo and adinono (2009), "Reduced Stall MIPS Architecture using Pre-Fetching Accelerator", International Conference on Electrical Engineering and Informatics, IEEE, ISBN: 978-1-4244-4913-2, pp. 611-616.
- Pejman lotfi, Ali-Asghar Salehpour, Amir-Mohammad Rahmani, Ali Afzali-kusha, and zainalabedin Navabi (2011), "dynamic power reduction of stalls in pipelined architecture processors", International journal of design, analysis and tools for circuits and sytems, Vol. 1, No. 1, pp. 9-15.
- Gautham P, Parthasarathy R, Karthi Balasubramanian (2009), "Low-power pipelined MIPS processor design", International symposium on integrated circuit (ISIC 2009), pp. 462-465.
- Pravin B. Ghewari, Mrs. Jaymala K. patil, Amit B. Chougule (2010), "Efficient hardware design and implementation of AES cryptosystem", International journal of engineering science and technology, Vol. 2(3), pp. 213-219.
- Rashmi S. Keote (2011), "Design of FPGA based Instruction Fetch & Decode Module of 32-bit RISC (MIPS) processor", 2011 International Conference on Communication Systems and Network Technologies, IEEE, ISBN: 978-0-7695-4437-3, pp. 409-413.
- Saeid Taherkhani, Enver Ever and Orhan Gemikonakli (2010), " Implementation of Non-pipelined and pipelined data encryption standard (DES) using Xilinx Virtex-6 Technology, 10[th] IEEE International Conference on computer and information technology(CIT 2010), pp. 1257-1262.